# Phonon-scattering-induced quantum linear magnetoresistance up to room temperature

Nannan Tang[1#], Shuai Li[2,3#], Yanzhao Liu[4], Jiayi Yang[1], Huakun Zuo[5], Gangjian Jin[5], Yi Ji[1], Bing Shen[1], Dingyong Zhong[1], Donghui Guo[1], Qizhong Zhu[6], Zhongbo Yan[1], Haizhou Lu[3,4], Jian Wang[7,8], Huichao Wang[1*]

[1]*Guangdong Provincial Key Laboratory of Magnetoelectric Physics and Devices, School of Physics, Sun Yet-sen University, Guangzhou, China.*

[2]*Hubei Engineering Research Center of Weak Magnetic-field Detection, College of Science, China Three Gorges University, Yichang, China.*

[3]*Department of Physics, Southern University of Science and Technology (SUSTech), Shenzhen, China.*

[4]*Quantum Science Center of Guangdong–Hong Kong–Macao Greater Bay Area (Guangdong), Shenzhen, China.*

[5]*Wuhan National High Magnetic Field Center and School of Physics, Huazhong University of Science and Technology, Wuhan, China.*

[6]*Guangdong Basic Research Center of Excellence for Structure and Fundamental Interactions of Matter, Guangdong Provincial Key Laboratory of Quantum Engineering and Quantum Materials, School of Physics, South China Normal University, Guangzhou, China.*

[7]*International Center for Quantum Materials, School of Physics, Peking University,*

*Beijing, China.*

[8]*Collaborative* Innovation *Center of Quantum Matter, Beijing, China.*

[#]These authors contribute equally: Nannan Tang, Shuai Li.

[*]e-mail: wanghch26@mail.sysu.edu.cn

**The realization of quantum transport effects at elevated temperatures has long intrigued researchers due to the implications for unveiling novel physics and developing quantum devices. In this work, we report remarkable quantum linear magnetoresistance (LMR) in the Weyl semiconductor tellurium at high temperatures of 40-300 K under strong magnetic fields up to 60 T. At high fields, the Weyl band features a large energy gap between the lowest and first Landau levels, which suppresses thermal excitation and preserves Landau quantization at high temperatures. The LMR is observed as long as majority carriers remain in the lowest Landau level without requiring monochromaticity, allowing it to persist up to room temperature. The inverse relationship between the LMR slope and temperature provides clear evidence that quantum LMR originates from high-temperature phonon scattering in the quantum limit, firstly demonstrating a theoretical prediction made nearly fifty years ago. This study highlights the key role of electron-phonon interaction and reveals an innovative quantum mechanism for achieving high-temperature LMR, fundamentally distinct from previous findings. Our results bridge a gap in the understanding of phonon-mediated**

**quantum-limit physics and establish strong magnetic fields at high temperatures as a promising platform for exploring novel quantum phenomena.**

Achieving quantum effect at room temperature has been one of the major pursuits in physics[1–6], crucial for advancing fundamental research and developing devices. However, macroscopic quantum transport effects are rarely observed at elevated temperatures and are typically limited to low temperature environments. Under strong magnetic fields, well resolved Landau quantization promotes discoveries in the quantum regime, especially when only the lowest few Landau levels are occupied[1,3–18]. These novel quantum phenomena emphasize predominant electron-impurity or electron-electron interactions at low temperatures. Notably, Abrikosov's quantum LMR arises from Coulomb-type impurities in the quantum limit[13], where carriers are in the lowest Landau level. At high temperatures, electron-phonon interaction becomes vital for transport properties. However, the understanding of possible behavior of solids in the quantum limit with dominant electron-phonon interaction remains largely unclear.

To preserve Landau quantization at high temperatures, a large Landau level gap is needed to minimize the effect of thermal excitation. Dirac and Weyl materials with unique energy dispersion that can harbor large Landau level gaps at large magnetic fields provide promising platforms. For example, room temperature quantum Hall effect was realized in graphene subject to a large magnetic field[1,6]. Elemental tellurium (Te) is a Weyl semiconductor that usually exhibits *p*-type doping in bulk crystals due to defects[10,19–21]. The Weyl feature of Te has been demonstrated by angle-resolved photoemission spectroscopy[22–24] and transport results including chiral anomaly effect and log-periodic quantum oscillations[10]. The system can easily reach low carrier density even at high temperatures[10,20,25–27], allowing the realization of quantum-limit regime under relatively low magnetic fields. Additionally, the relativistic Weyl fermions have a noticeable effect on transport even when the Fermi level is near the top of the valence band[10,27,28]. These characteristics make the system attracting for exploring quantum-limit effects at elevated temperatures, but high-temperature studies are sparse and limited to low magnetic fields.

Through systematic electron transport measurements of the Weyl semiconductor Te at high temperatures and magnetic fields, this work demonstrates a mechanism for achieving high-temperature quantum LMR by harnessing electron-phonon scattering in the quantum limit. In self-doped Te crystals with a low carrier density of ~$10^{17}$ cm$^{-3}$, the LMR shows remarkable linearity and robustness across a broad magnetic field range up to 60 T and a wide temperature range of 40-300 K. Because of the Weyl band structure, the large Landau level separation at high fields enables most carriers to occupy the lowest Landau level even at room temperature, preserving the characteristic transport behavior of the quantum limit. Quantitatively, both the transverse and longitudinal LMR display slopes inversely dependent on temperature, aligning with theoretical derivations. These results provide the first experimental confirmation of quantum LMR induced by phonon scattering. This discovery hints at intriguing high-temperature quantum phenomenon at strong magnetic fields.

**Dominated phonon scattering at high temperatures**

We synthesized Te single crystals using methods detailed in our prior research (see Methods)[20]. The X-ray diffraction results of the sample (Extended Data Fig. 1a) match well with the standard pattern of trigonal Te (JCPDS No. 36–1452). The energy dispersive spectroscopy shows that the crystal is pure, with only Te element detected (Extended Data Fig. 1b). With the chiral crystal structure (Fig. 1a), the *p*-type trigonal Te possesses a simple band structure near the Fermi level, featuring two hole bands that cross at the Weyl points[10,28] (Fig. 1b). The samples studied are long and narrow (inset of Fig. 1c), with lengths ranging from 0.1 to 20 mm, widths between 50 to 250 μm, and uniform thicknesses between 50 to 200 μm.

Figure 1c shows the measured resistivity-temperature characteristics of a typical Te crystal, which illustrate its doped semiconductor nature (see Methods and Extended Data Fig. 2). The Hall results in Fig. 1d exhibit weak nonlinear behavior at low magnetic fields, a feature commonly seen in Te crystals. This can be attributed to the influence of an impurity band created by Te vacancies due to the absence of degenerate valence bands[19,21,27]. Using two-carrier analyses (see Methods and Extended Data Fig. 3), we obtain the carrier density ($n$) and mobility ($\mu$) of holes from the valence band, as shown in Figs. 1e and 1f. The increase of mobility at low temperatures is a typical behavior in semiconductors resulting from impurity scattering[29,30]. In this regime, carriers are scattered by the electrostatic potential of impurities, affecting their mobility. As temperature rises, carriers gain more kinetic energy and spend less time near impurities, leading to a mobility change that follows $\mu \sim T^{3/2}$ in doped semiconductors[29,30]. The quantitative fit (blue dashed line in Fig. 1f) is well consistent with the experimental data, revealing dominated impurity scattering at low temperatures. As temperature further increases, the lattice vibration scatterings become more significant, resulting in the decrease of mobility. The red dashed line in Fig. 1f represents a fit of $\mu \sim T^{-3/2}$ which is the characteristic of phonon scattering[29,30]. We note a deviation from this trend in the intermedium temperature range of 40 - 100 K, which can be attributed to additional scattering mechanisms[31]. By considering both impurity and phonon scattering, the temperature dependence of mobility above 40 K is accurately reproduced (pink curve in Fig. 1d). According to the fitting parameters, the coefficient of the phonon scattering term ($C_p \sim 388.64$) is much larger than that of impurity scattering ($C_i \sim 0.0038$) (Extended Data Fig. 4), which highlights electron-phonon scattering as the main factor affecting transport properties at temperatures above 40 K.

**High-temperature LMR at large magnetic fields**

The isothermal MR behavior of the Te crystal is shown in Figs. 2a and 2b, which exhibits a pronounced temperature dependence in response to the applied magnetic field. Consistent with previous study, weak log-periodic oscillations are observed in the MR at low temperatures (Extended Data Fig. 5), which are attributed to discrete scale invariant quasi-bound states formed by massless Weyl fermions[8,9,32,33]. The low-temperature MR shows an approximately quadratic dependence on the magnetic field. As the temperature increases, the MR shows a quadratic dependence at low magnetic fields while the high-field MR transitions to a linear dependence, and this remarkable LMR persists up to room temperature (Fig. 2c). To quantitatively characterize the MR, we fit the high-field data using the power-law expression MR~$B^m$. The temperature dependence of the fitting parameter $m$ is displayed in Fig. 2d. Notably, at high fields, the exponent $m$ remains nearly 1 for temperatures above 40 K, signifying nearly ideal LMR behavior. Furthermore, the first derivative d$R$/d$B$ of the MR data remains almost constant across high magnetic fields (Fig. 2e), providing additional confirmation of the LMR feature. Based on these results, it is evident that both elevated temperature and strong magnetic fields are essential for the emergence of LMR in Te crystals.

The comparison between Figs. 1f and 2d reveals that the LMR appears in a relatively high-temperature regime where phonon scattering dominates. This suggests a connection between MR characteristics and temperature-dependent scattering processes. Under strong magnetic fields, Landau quantization becomes important, particularly when carrier scattering within the lowest Landau level plays a key role[1,6,–12,16]. The low carrier density of the sample indicates a low quantum-limit field $B_Q$, i.e., at which the system enters the quantum limit that all carriers occupy the lowest Landau level. For low-density materials without observable Shubnikov–de Haas (SdH) oscillations, $B_Q$ is typically determined by the carrier density $n$. The relationship is given by $B_Q = (2\pi^4)^{1/3}\hbar n^{2/3}/e \approx 3.8\times10^{-11} n^{2/3}$. Using high-density samples that show clear SdH oscillations, we confirm that this formula accurately identifies the quantum limit of our material (see Methods). For the low-density samples displaying LMR, we estimate $B_Q$ based on the hole density in the valence band (Fig. 1e). For comparison, the crossover field $B_c$ for the onset of LMR is shown in Fig. 2f together, which is determined from Fig. 2e by the deviation from the linear fit of d$R$/d$B$ at high fields as indicated by arrows. It is noted that $B_c$ and $B_Q$ are very close with $B_c$ slightly larger than $B_Q$ at high temperatures above 200 K. This indicates that the high-temperature LMR occurs within the quantum-limit regime (see the effect of thermal excitation in Discussion).

Quantum-limit LMR is often attributed to the mechanism proposed by Abrikosov[13]. In his original theory, scattering from screened Coulomb impurities leads to quantum LMR in systems with a linear energy spectrum[13]. This mechanism was later extended to systems with quadratic dispersion, as well as those with long-range Gaussian potential impurity[14]. However, an insurmountable obstacle arises when attempting to reconcile our results with this mechanism. The impurity scattering dominates below 40 K in Te crystals, but the LMR appears only in the high-temperature regime where phonon scattering dominates. This requirement for relatively high temperatures also conflicts with many other classical models based on factors such as carrier density fluctuations[34], crystal thickness variations[35], doping inhomogeneity[36], or current distortion[37]. These mechanisms rely on inherent sample features and typically produce more pronounced LMR at low temperatures. Control experiments using both four-probe and two-probe setups show that the LMR is independent of the electrode configuration, indicating that it is not related to measurement positions or current paths. Furthermore, classical or semiclassical mechanisms apply when Landau levels are not well resolved, and LMR emerges from low magnetic fields typical of of the classical regime. This contrasts with our findings, where LMR is observed only at high fields in the quantum-limit regime. Further details on excluding these mechanisms are provided in Supplementary information.

**Phonon-scattering-induced quantum LMR**

In our study, carrier scattering primarily arises from impurities and phonons. The total scattering time, $\tau$, is described by $1/\tau = 1/\tau_i + 1/\tau_p$, where $\tau_i$ and $\tau_p$ represent the scattering times due to impurities and phonons, respectively. As the temperature increases, the scattering source gradually shifts from impurity dominated ($\tau_i \ll \tau_p$) to phonon dominated ($\tau_p \ll \tau_i$), as shown in Fig. 1f. The corresponding MR displays a nearly quadratic dependence on the magnetic field at low temperatures, and transitions to a linear dependence at elevated temperatures. Therefore, to understand the observed LMR, an investigation into electron-phonon scattering is indispensable.

Under Landau quantization, the three-dimensional wavevector of the system is transformed: $k_\parallel$ parallel to the field remains unchanged; $k_\perp$ perpendicular to the field is replaced by the Landau index $\nu$, and a degeneracy index that labels the degenerate Landau states. As a result, carriers scattered by phonons transition between states characterized by $k_\parallel$, $\nu$ and the degeneracy index. Utilizing a rigorous quantum framework, we calculated the MR by incorporating electron-phonon scattering into the scattering matrix (see Supplementary information). It is worth noting that the quantum LMR proposed by Abrikosov[13], as well as LMR attributed to other types of impurity[14], were derived using this same theoretical approach but with different scattering potentials. When a large number of Landau levels are occupied, the phonon-scattering dominated MR exhibits a quadratic dependence on the magnetic field[38,39]. As the magnetic field increases and fewer Landau levels are occupied, the possible scattering processes between states with different $\nu$ are greatly reduced. In the quantum limit, the scattering process becomes much simpler, occurring within a highly degenerate one-dimensional band. Specifically, scattering is limited to states where the change in $k_\parallel$ is either zero or matches the difference between the Fermi wavevectors, as depicted in Fig. 3a. This type of scattering leads to a distinctive field dependence of the scattering matrix, which gives rise to the LMR,

$$\rho_\perp/\rho_0 = \frac{2e\hbar}{3\pi m^*} \frac{B}{k_B T} \left[ \ln\left(\frac{3\sqrt{\pi} k_B T \tau}{2\hbar}\right)^2 - \gamma \right], \qquad (1)$$

where $\rho_\perp$ is the transverse magnetoresistivity, $\rho_0$ is the zero-field resistivity, $e$ is the elementary charge, $\hbar$ is the reduced Planck constant, $m^*$ is the effective mass, $k_B$ is the Boltzmann constant, $\tau$ is the average electron-phonon scattering time at zero field, and $\gamma = 0.577$ is the Euler constant. A special characteristic of the phonon scattering is that, the longitudinal magnetoresistivity $\rho_\parallel$ is also linear in this regime,

$$\rho_\parallel/\rho_0 = \frac{e\hbar}{3m^*}\frac{B}{k_B T}. \qquad (2)$$

Our theoretical results align with the theory of phonon-scattering-induced LMR in the quantum limit, proposed nearly fifty years ago[38], which has not previously been demonstrated experimentally.

To investigate the relationship between the experimental phenomena and this mechanism, we further examine the field orientation dependence of the LMR. In the transverse case when the field is perpendicular to the (10$\bar{1}$0) plane, LMR is sustained over a wide range of magnetic fields (Fig. 3b), consistent with the results shown in Fig. 2b. In the longitudinal case when the field is parallel to the current along the [0001] direction, the behavior at lower magnetic fields is affected by additional effects, such as the chiral anomaly[10], which are beyond the scope of this article. Nevertheless, at higher magnetic fields, LMR is also observed in the longitudinal configuration (Fig. 3c). Therefore, both transverse and longitudinal MR exhibit a linear dependence on the magnetic field, which is qualitatively consistent with the theoretical predictions.

The slopes of $\rho_\perp/\rho_0$ and $\rho_{//}/\rho_0$ measured at different temperatures are shown in Figs. 3d-

f. The quantitative fitting of the transverse data for sample A8 yields $m^* = 0.17 \pm 0.02$ $m_e$ and $\tau = (7.2 \pm 0.5) \times 10^{-13}$ s with $m_e$ is the free electron mass. The fit for the transverse data of sample A2 gives consistent values, with $m^* = 0.18 \pm 0.02\ m_e$ and $\tau = (1.01 \pm 0.08) \times 10^{-12}$ s, and the temperature dependence of longitudinal MR in A2 corresponds to the same effective mass $m^* = 0.18 \pm 0.01\ m_e$. Consistent results are also obtained in other samples (Extended Data Fig. 6), and they agree well with previous reports of low-carrier-density Te crystals, which show $m^* \approx 0.19\ m_e$ and $\tau \approx 1.0 \times 10^{-12}$ s[20,26]. Additionally, the electron-phonon scattering time $\tau$ can be independently verified using the relationship $\tau^{-1} = 3(2m^* k_B T)^{3/2} E_d^2 / 8\pi^{1/2} D v_s^2 \hbar^4$ (see Supplementary Information), where the average sound velocity $v_s = 2600$ m/s, the material density $D$ is $6.237 \times 10^3$ kg/m$^3$, and the deformation potential constant $E_d$ is 6.5 eV for tellurium[40–42]. With these parameters and an effective mass of $m^* = 0.18\ m_e$, the calculated scattering time is $\tau = 2.0 \times 10^{-12}$ s, which is consistent with the values derived from the LMR fits. It is noted that, for $\tau \approx 10^{-12}$ s, Eq. (1) is dominated by the linear term rather than the logarithmic term. The temperature and angular dependences together provide evidences for the high-temperature LMR originating from phonon scattering in the quantum limit, well demonstrating the long-standing theoretical prediction.

**Discussion**

It is important to note that the phonon-scattering-induced LMR occurs when the low-density carriers are confined to the lowest Landau level and the Landau level gap is larger than the thermal excitation energy. This condition is difficult for quadratically dispersive materials at high temperatures, especially room temperature. However, Dirac and Weyl materials can have large Landau level gaps because of their unique energy dispersions[1,6]. Our evaluation of the Landau level structure in the Weyl semiconductor Te reveals large energy gaps between Landau levels (see Section IIA in Supplementary information). Because the carrier density is low, holes are squeezed to the band edge at large magnetic fields. At room temperature, the estimated minimum energy required for carriers at the edge of the lowest Landau band to reach the next band is at least 39 meV at 15 T. At higher magnetic fields up to 60 T—where our main results are focused—the Landau level gaps significantly exceed the thermal excitation energy $k_\text{B}T$ (Fig. 4b). Additionally, our tellurium samples show a mobility of about 1600 cm$^2$V$^{-1}$s$^{-1}$ at room temperature, which is larger than that of many other materials and indicates smaller broadening effect[43-45]. After evaluating all possible scenarios (see Section IIB Supplementary information), we find that the broadening energy is less than 3.8 meV, much smaller than the Landau level gap. Therefore, the Landau quantization remains robust in our system even at room temperature.

We further investigate carrier occupation to clarify the influence of thermal excitation. At elevated temperatures and relatively lower fields, a "tail" develops in the carrier distribution, allowing some carriers to occupy higher Landau levels beyond the lowest one. On the other hand, the increase of Fermi energy to maintain a fixed carrier density[29,30] (Fig. 4c) restricts hole occupation in higher Landau levels. We calculate the density of states and spectral function, incorporating the Fermi-Dirac distribution function, which demonstrates that the lowest Landau level remains predominantly occupied even at room temperature (Figs. 4d and 4e). Specifically, numerical results indicate that 70.65% of carriers remain in the lowest Landau level at 20 T and 300 K, and 63.13% of carriers are still in this band even at 15 T. Considering the limitations inherent to the low-energy model (see Supplementary information), this value should be larger. Under these conditions, the MR is expected to remain linear or quasi-linear as the majority of carriers contribute from the lowest Landau level (I in Fig. 4f). The presence of additional scattering channels between Landau levels (II and III in Fig. 4f) causes the MR to deviate from perfect linearity, but is far from quadratic relationship that would require significant occupation of multiple Landau levels[38,39] (Fig. 4g). It is noted that the simultaneous occupation of lowest few Landau levels can hinder the observation of phenomena which rely on abrupt changes in carrier occupation with varied magnetic field, such as the SdH effect. However, the LMR persists as long as the majority carriers occupy the lowest Landau level without requiring monochromaticity, allowing it to survive up to room temperature.

We elucidate that several conditions are crucial for observing phonon-scattering-induced quantum LMR, all of which are ideally satisfied in the studied Weyl semiconductor tellurium. First, the band structure near the Fermi surface should be simple, preferably involving only a single band, to avoid complexities from multi-band contributions. Second, the manifestation of LMR requires the system to be in the quantum limit, or at least for the majority of carriers to occupy the lowest Landau level, over a broad range of magnetic fields. This demands a relatively low carrier density and the application of a strong magnetic field. Most importantly, phonon scattering must dominate over impurity scattering. Thus, this mechanism exhibits unique characteristics different from previously known models and allows the observation of LMR at high temperatures, where phonon scattering prevails. Furthermore, both transverse and longitudinal MR show the linear dependence.

The discovery of phonon-scattering-induced quantum LMR enhances the ability to manipulate LMR under ambient conditions (Supplementary Table 3). Though LMR was previously realized at high temperatures through classical mechanisms related to inhomogeneity[35,46], its occurrence depending on sample quality is elusive, which is unfavorable for the control. The room-temperature quantum LMR reported in Dirac material graphene and $Cd_3As_2$ microribbons has not been understood regarding the mechanism[5,36,46]. Our demonstration of room-temperature quantum LMR in bulk Te offers advantages compared to the micro- and nano-scale systems, which require intricate fabrication and precise gate tuning to achieve comparable performance. Bulk Te single crystals are more stable and accessible, showing superior linearity and reproducibility. More importantly, the observation of quantum LMR induced by phonon scattering advances the understanding of high temperature quantum-limit effects. To date, the intriguing phenomena emerging in the quantum limit of solids are confined to stem from electron-impurity or electron-electron interactions[7–12]. The novel effect of electron-phonon interaction at high temperatures remains to be explored[6]. By demonstrating quantum LMR in the Weyl semiconductor Te as a phonon-mediated phenomenon, we show that Dirac and Weyl systems under strong magnetic fields offer an appealing platform for future investigations.

**Methods**

**Sample synthesis**

The low-carrier-density Te single crystals (A-type) were grown in a two-step process. First, Te powder (4.13 g, purity 99.99%) and Hf powder (1.04 g) were sealed in a quartz tube under a high-purity argon atmosphere and sintered in a muffle furnace. The mixture was gradually heated up from room temperature to 1000 °C over 13 h, held at the temperature for 5 h, and then cooled naturally to room temperature. In the second step, the resulting precursor was mixed with $I_2$ (0.51 g), vacuum-sealed in a vapor transport tube, and subjected to a temperature gradient: the high-temperature side was heated to 500 °C and the low-temperature side to 400 °C over 10 h. This temperature difference was maintained for 350 h before cooling naturally to room temperature. The final products consisted mainly of long, narrow Te single crystals and a small amount of $HfTe_5$ with a layered structure. By reducing the amount of Te powder while keeping other conditions unchanged, higher-carrier-density Te crystals, identified by the Hall results and the presence of SdH oscillations, were obtained.

**Characterization and transport measurements**

The X-ray diffraction patterns were measured using a PANalytical Empyrean 2 diffractometer with a wavelength of 1.54 Å X-ray excited by a copper target. The energy dispersive spectra were collected with a BRUKER X-ray energy spectrometer on a Zeiss EVO MA10 scanning electron microscope. The transport experiments were performed in a 14T-PPMS (Physical Property Measurement System, Quantum Design) and a pulsed high magnetic field facility at the Wuhan National High Magnetic Field Center. To ensure good ohmic contact, 10 nm thick Pd and 50 nm thick Au were deposited on the samples as electrodes by thermal evaporation. Silver epoxy was then used to attach the gold wires (diameter 25 µm) to the Pd/Au electrodes. The standard four-probe method was used to measure MR, and the six-probe method enabled simultaneous measurement of Hall and MR signals. The excitation current was always applied along the [0001] direction. The perpendicular field was oriented along the normal direction of the ($1\bar{2}10$) plane.

**Doped semiconducting behavior of Te crystals**

In a doped-semiconductor, as the temperature increases, the system undergoes three distinct regimes: weak ionization, saturated ionization, and intrinsic excitation. As shown in Extended Data Fig. 2a, the zero-field resistivity curve of A-type Te crystals exhibits typical semiconducting behavior over the temperature range of 2-390 K. Above 330 K, the resistivity decreases with rising temperature due to an increase in carrier density from intrinsic excitation. By analyzing the data using an Arrhenius plot[47] ($\ln\rho \propto E_g/2k_BT$, where $k_B$ is the Boltzmann's constant), we determine an energy band gap $E_g$ of approximately 0.2 eV (Extended Data Fig. 2b), consistent with previous studies. Between 40 and 330 K, the system is in the saturated ionization regime, where impurity acceptors are thermally activated and the overall carrier density remains nearly constant.

Below 40 K, the resistivity shows an obvious increase as the temperature is lowered. In this low-temperature range, the system is in the weak ionization regime and the carrier density decreases due to the freeze-out effect.

**Hall analyses**

In our sample, the Hall trace exhibits nonlinear behavior at low magnetic fields (Fig.3a), which is usually attributed to the presence of an impurity band. We use a two-carrier model to fit the Hall conductivity tensor $\sigma_{xy} = \frac{\rho_{yx}}{\rho_{xx}^2 + \rho_{yx}^2} = \left[ \frac{n_1 \mu_1^2}{1+(\mu_1 B)^2} + \frac{n_2 \mu_2^2}{1+(\mu_2 B)^2} \right] eB$ (Extended Data Fig. 3b). The resulting carrier densities and mobilities at different temperatures are shown in Extended Data Figs. 3c and 3d. The data indicate that the samples have high-mobility holes ($n_v$) from the valence band and low-mobility holes ($n_i$) from the impurity band. When the impurity band is located near the valence band but just above its top, electrons in the impurity band can easily transition to the valence band with low energy at low temperatures, creating holes. As a result, the hole density in the impurity band is relatively high at low temperatures, while it is lower in the valence band. As the temperature increases, more electrons from the valence band can be thermally excited to the impurity band, leading to an increase in the hole density in the valence band and a decrease in the hole density in the impurity band. The temperature dependence of the carrier density can thus be well understood by considering the influence of the impurity band. The observed LMR phenomenon is closely related to holes in the valence band, which forms quantized Landau levels under strong magnetic fields.

In tellurium, due to the absence of degenerate valence bands[19,21,27], it is common to use the linear slope at high magnetic fields for carrier analyses. For comparison, Extended Data Figs. 3e and 3f show the results obtained using the high-field linear slope. It can be seen that the carrier density obtained using the high-field linear slope is slightly larger than the total carrier density derived from the two-carrier model, but overall the values are close showing consistency between the two approaches.

**Weyl physics and log-periodic oscillations**

The studied Te sample has a relatively low carrier density, making it an ideal platform for studying transport properties beyond the quantum limit. The critical field at which the system reaches the quantum limit can be estimated by the formula[48]: $B_Q = (2\pi^4)^{1/3}\hbar n^{2/3}/e \approx 3.8\times10^{-11}\, n^{2/3}$, which $n$ is the carrier density, $\hbar$ is the reduced Planck constant, and $e$ is the elementary charge. For A-Type sample with a carrier density of ~$4.28\times10^{16}$ cm$^{-3}$ at 1.7 K, the estimated $B_Q$ is 4.66 T. As shown in Fig. 2a and Extended Data Fig. 5a, the MR curves at 1.7 K in a transverse field configuration display oscillations even above 4.66 T, which cannot be explained by the SdH effect.

Using standard analysis methods, the oscillatory components become clearer by subtracting a appropriate smooth background or employing the second derivative. The nearly identical oscillation patterns and periodicity obtained from different methods (Extended Data Fig. 5b) confirm the intrinsic nature of these quantum oscillations. Moreover, the oscillation peaks and dips exhibit a logarithmic dependence on magnetic field, as illustrated in Extended Data Fig. 5c, where peaks and dips are assigned to integers and half-integers, respectively. The linear relationship between $B_i$ and $i$ in a semi-log scale demonstrates logarithmic periodicity. Extended Data Fig. 5d displays the Hall data collected simultaneously. The presence of log-periodic oscillations is further confirmed by analyses (Extended Data Figs. 5e and 5f).

**Quantum limit of the samples**

The identification of the quantum-limit field $B_Q$ in low-density materials without SdH effect is usually based on the carrier density $n$ using the formula $B_Q = (2\pi^4)^{1/3}\hbar n^{2/3}/e$. We use Te crystals with higher carrier density that display the SdH effect to verify this formula. In these samples, the SdH effect is observed in the low-temperature MR (Supplementary Information). Fast Fourier transform (FFT) analysis reveals an oscillation frequency peak at 27 T, indicating a quantum limit field of about 27 T. Substituting the carrier density of the sample ($\sim 5.6\times10^{17}$ cm$^{-3}$) into the theoretical formula yields a calculated quantum limit field of 25.88 T, in agreement with the SdH oscillation results and confirming the reliability of the calculation method. For sample A8 with a carrier density of $\sim 4.28\times10^{16}$ cm$^{-3}$ at 1.7 K, the estimated $B_Q$ is 4.66 T. Taking into account the correction factor due to Fermi surface anisotropy (27/25.88), the quantum-limit field at 1.7 K is determined to be 4.86 T.

The estimation is further confirmed by an alternative quantitative calculation method[20,49]. For tellurium with a low carrier density, the Fermi surface consists of a single ellipsoid arising from spin-split energy bands, with a spin degeneracy $g_s$=1 and a valley degeneracy $g_v$=2. Using the formula $n=g_s g_v (2eF/\hbar)^{3/2}/6\pi^2$, we calculated a frequency $F$ of approximately 3.85 T. This calculation assumes the Fermi surface is isotropic, which introduces some deviation from the actual situation. To estimate the anisotropy, we analyzed SdH oscillations measured in different directions. In the perpendicular configuration, where the magnetic field is aligned along the [10$\bar{1}$0] direction and the current $I$ flows along [0001] direction, the FFT spectrum reveals a frequency peak $F_\perp$ (Extended Data Figs. 7a and 7c). According to the relation $F = \hbar S_F / 2\pi e$, the projection area of the Fermi surface, denoted as $S_{F\perp}$, can be determined. In the longitudinal case, where both the magnetic field and the current lie within the plane along the [0001] direction, the FFT result shows two frequency peaks (Extended Data Figs. 7b and 7d), corresponding to two Fermi surface cross-sections. As illustrated in Extended Data Figs. 7e and 7f, the smaller peak represents the overlapping region of two ellipsoids, so we use the larger one, denoted as $S_{F//}$, to estimate the anisotropy of the ellipsoid. It is found that $S_{F\perp}$ is about twice that of $S_{F//}$ in all the samples with observable SdH oscillations. According to the anisotropy, the frequency $F_\perp$ in the perpendicular field is corrected to be about 4.87 T. This value is very close to the above estimated $B_Q \sim 4.86$ T. Therefore, the quantum limit field is nearly equal to, or slightly lower than, the crossover field to LMR, indicating a strong connection between the pronounced LMR behavior and the quantum limit regime.

**Data availability** Data measured or analyzed during this study are available from the corresponding author on reasonable request.

**Acknowledgements**

We thank M. Lu, H. Liu, R. Q. Wang, and M. X. Deng for discussions. We acknowledge the support from the National Natural Science Foundation of China (grant nos. 21BAA01133, 12374052, 92165204, 12488201, 12350402, 11925402), the National Key Research and Development Program of China (grant no. 2022YFA1403700, 2022YFA1405304), Guangdong Provincial Key Laboratory of Magnetoelectric Physics and Devices (grant no. 2022B1212010008), Guangdong Basic and Applied Basic Research Foundation (grant nos. 2023A1515010487, 2023B0303000011), Hubei Provincial Natural Science Foundation of China (grant no. 2024AFB289), Guangdong Provincial Quantum Science Strategic Initiative (grant nos. GDZX2401009, GDZX2401002), Guangzhou Basic and Applied Basic Research Foundation (grant no. 2025A04J5405) and Guangdong province (grant no. 2020KCXTD001). The experiments are supported by the Guangdong Provincial Key Laboratory of Magnetoelectric Physics and Devices. The numerical calculations are supported by Center for Computational Science and Engineering of SUSTech.

**Author contributions**

N.T. and S.L. contributed equally to this work. H.W. conceived and supervised the research. N.T., Y.J. and B.S. grew the samples; N.T., J.Y., G.J., H.Z. and Y.J. carried out the experiments; S.L. performed calculation. H.W., S.L., J.W. and H.L. proposed the theoretical model; H.W., N.T., S.L., Y.L., J.Y., H.L. and J.W. analyzed the data; Q.Z., Z.Y., D.Z. and D.G. participated in the correlated discussion. H.W., N.T. and S.L. wrote the manuscript, and all authors commented on the manuscript.

**Competing interests**

The authors declare no competing interests.

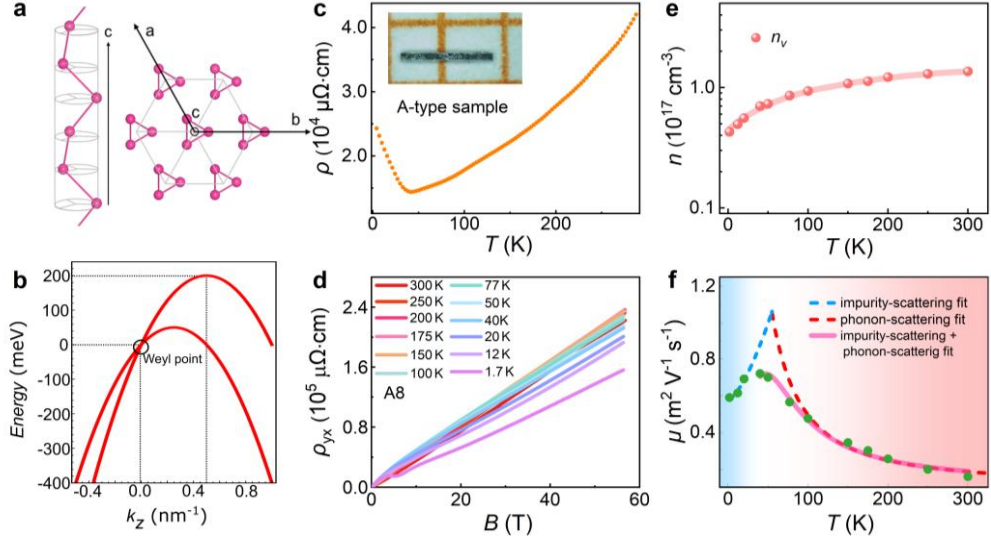

**Fig. 1 | Doped-semiconducting behavior of the Weyl semiconductor Te. a,** Crystal structure of trigonal Te. **b,** Electronic band structure of tellurium (k-path L-H-A) plotted by the eigenvalues of an effective low-energy model (see Section II in Supplementary information). **c,** Typical resistivity versus temperature curve for an A-type Te crystal. (Inset) Optical image of the Te bulk crystal. **d,** Hall resistance at different temperatures. The weak nonlinear behavior at low magnetic fields is typically attributed to the influence of an impurity band created by vacancies. **e,** Carrier density in the valence band as a function of temperature obtained using a two-carrier model. **f,** Carrier mobility as a function of temperature. The increase in mobility at low temperatures is due to impurity-scattering, which leads to $\mu \sim T^{3/2}$ (blue dashed line). The mobility change caused by phonon scattering follows $T^{-3/2}$ (red dashed line). By considering both impurity and phonon scattering effects (pink dashed line), the temperature dependence of mobility above 40 K is accurately described, with the much larger coefficient of the phonon scattering term indicating it is the dominant factor affecting transport properties at temperatures above 40 K.

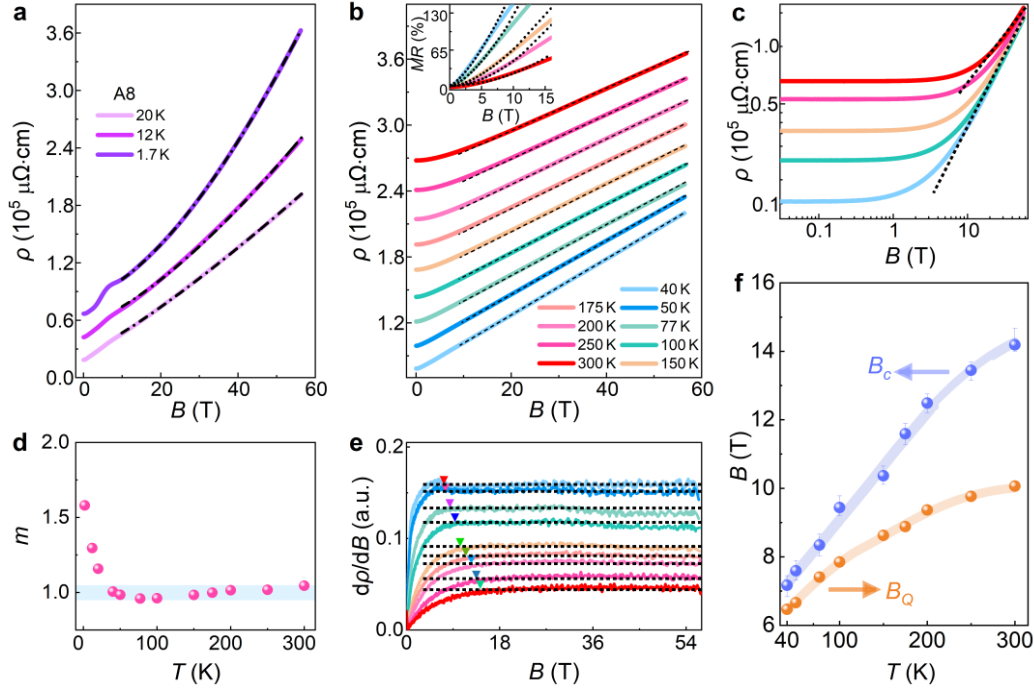

**Fig. 2 | LMR at higher temperatures and magnetic fields (sample A8). a,** The low-temperature MR shows a nearly quadratic dependence on the magnetic field. **b,** As the temperature rises above 40 K, the MR at high fields evolves into a linear dependence on the magnetic field. For clarity, data curves are shifted. (Inset) The MR shows a quadratic dependence at low magnetic fields. **c,** The MR curves above 40 K in a log-log scale illustrate the deviation from linearity at low fields. **d,** Temperature dependence of the exponent $m$ obtained by fitting of MR data (above 10 T) using MR~ $B^m$. **e,** First derivatives of the MR. The arrow refers to the crossover field $B_c$ where the MR slope changes from the high-field constant, marking the point where MR begins to deviate from the large-field linear dependence. The black dashed lines are guides for eyes. **f,** The estimated quantum limit field $B_Q$ and the crossover field $B_c$ to LMR at different temperatures. The smaller $B_Q$ compared to $B_c$ indicates the pronounced LMR behavior is established in the quantum limit regime.

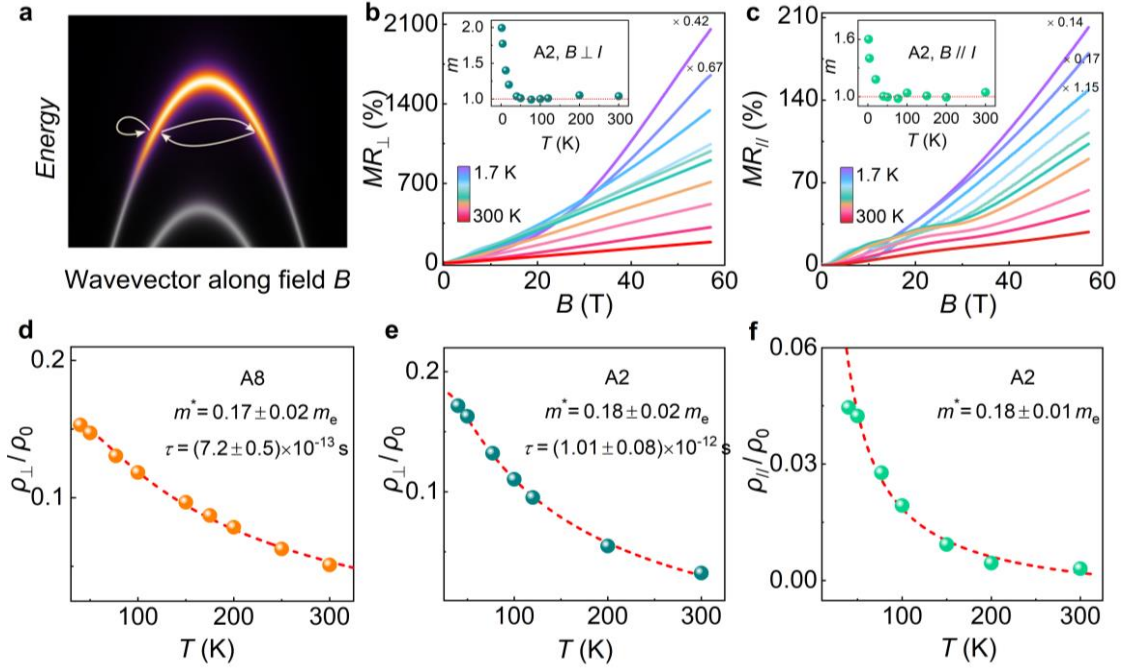

**Fig. 3 | Phonon-scattering-induced quantum LMR. a,** Schematic of the two primary scattering channels (arrows) responsible for phonon-scattering-induced LMR when carriers are confined to the lowest Landau level. **b,** MR behavior of sample A2 in the transverse magnetic field configuration. **c,** MR in the longitudinal configuration with the magnetic and electric fields aligned along the *c* axis. Insets show the exponent *m* at different temperatures. For clarity, lower-temperature data are scaled down. **d-e,** The transverse and **f,** longitudinal LMR slopes show an inverse dependence on temperature. The red dashed lines represent fits using the phonon-scattering-induced LMR formulas, Equations (1) and (2). The fitted effective mass is about $0.16 \pm 0.04$ $m_e$ for sample A1 and $0.18 \pm 0.01$ $m_e$ for sample A2. The fitted scattering time is around $\tau = 1.0 \times 10^{-12}$ s. These parameters are consistent with our calculations and previous reports for the low-carrier-density tellurium.

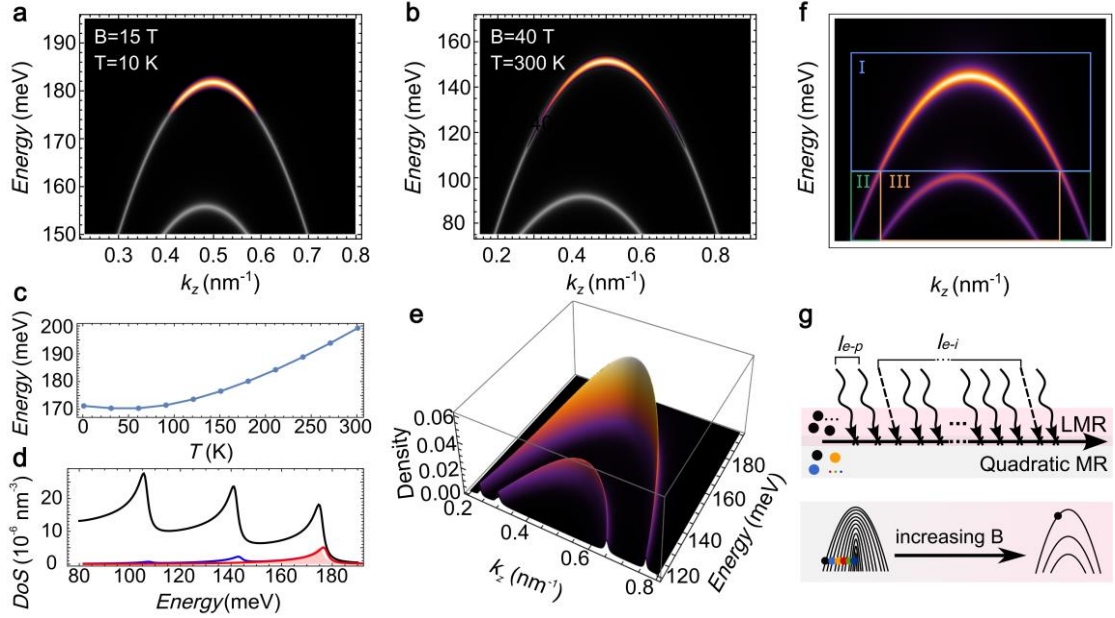

**Fig. 4 | LMR arising from dominated carriers in the lowest Landau level. a**, Carrier occupation at low magnetic fields and temperatures. **b**, Carrier occupation at high fields and temperatures. The colored and gray region denote holes and electrons, respectively. **c**, Fermi energy as a function of temperature. **d**, Density of states (black curve), total carrier occupation (blue curve), and carrier occupation in the lowest Landau level (red curve with shading). **e**, Hole carrier occupation in the lowest and first Landau levels. The data in d and e are calculated at 20 T and 300 K, revealing the lowest Landau level remains predominantly occupied. **f**, Transport properties arise from three primary contributions when more than the lowest Landau level is occupied. **g**, Schematic illustration of phonon-scattering-induced MR. Colored dots indicate carriers with different wavevectors along the magnetic field direction. Wavy and dashed lines represent phonon and impurity scattering, respectively. The mean free path for electron-phonon (e-p) scattering is much shorter than for electron-impurity (e-i) scattering. The phonon scattering in the quantum limit leads to LMR.

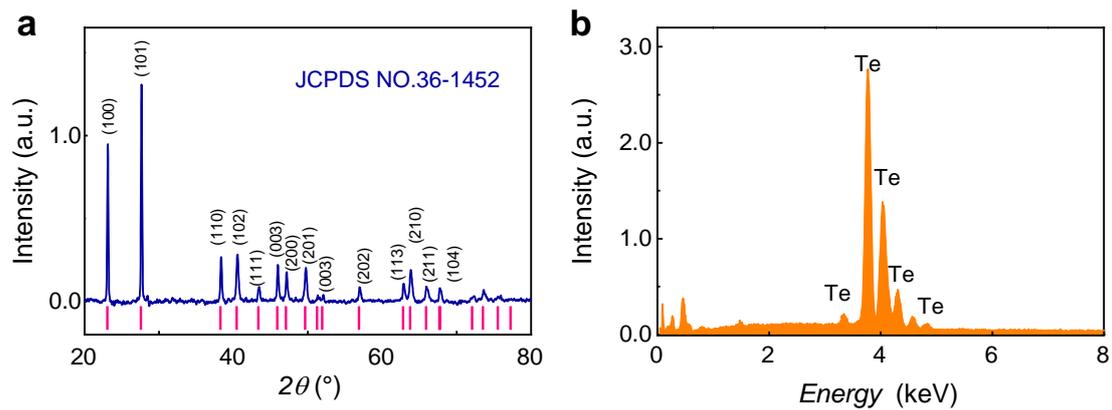

**Extended Data Fig. 1 | Characterization of Te crystals. a,** The powder X-ray diffraction pattern illustrates a crystal structure consistent with trigonal Te. **b**, Energy dispersive spectroscopy shows that no other elements are detected in the pure Te crystal.

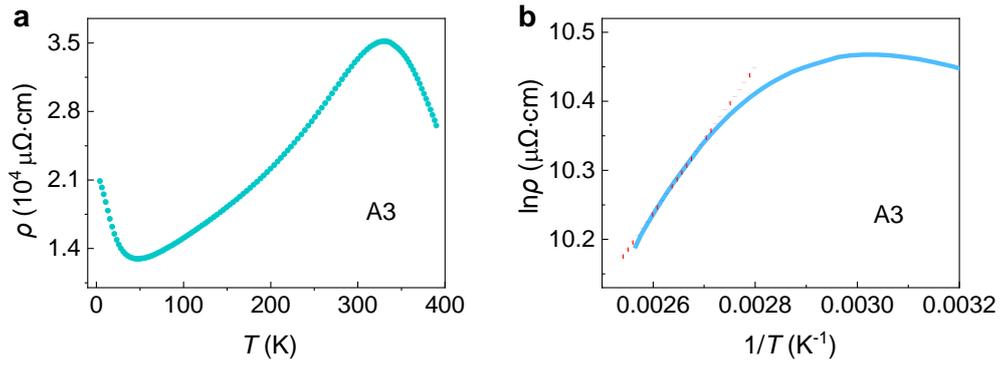

**Extended Data Fig. 2 | Doped-semiconducting feature of an A-type Te crystal. a**, Resistivity versus temperature curve. **b,** Arrhenius fit of the resistivity data from 360 to 390 K using the relation $\ln\rho \propto E_g/2k_BT$, where $k_B$ is Boltzmann's constant[47]. The extracted energy band gap $E_g$ is around 0.2 eV, consistent with previous report[48].

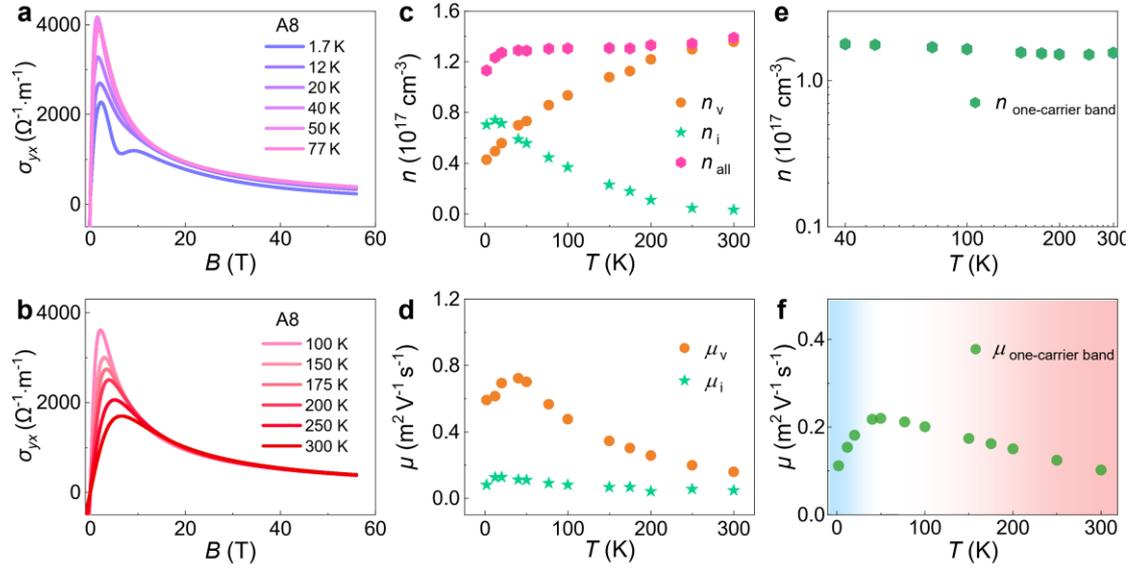

**Extended Data Fig. 3 | Comparison of Hall analyses results using two-carrier and single-carrier models. a**, **b**, Hall conductivity $\sigma_{yx}$, **c,** Carrier densities and **d,** mobilities at various temperatures obtained from fitting $\sigma_{yx}$. The total carrier density remains around $1.30\times10^{17}$ cm$^{-3}$ across the temperature range of 40-300 K. **e,** Carrier density and **f,** mobility as functions of temperature extracted from the high-field Hall resistivity. The total carrier density is about $1.60\times10^{17}$ cm$^{-3}$ above 40 K.

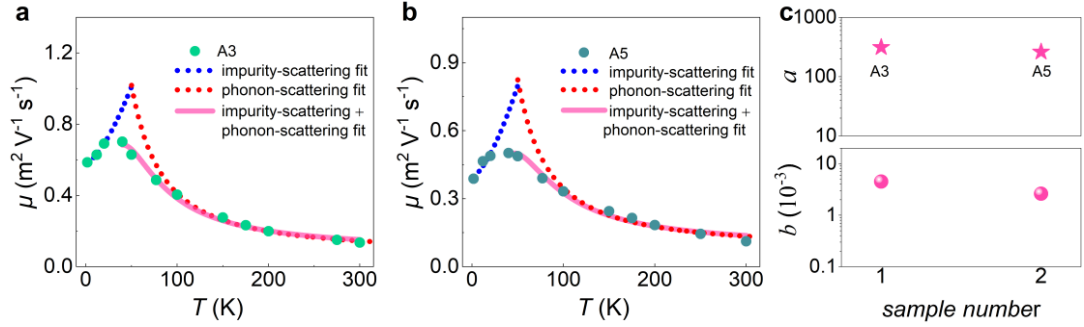

**Extended Data Fig. 4 |** Analysis of mobility change in other samples. **a**, **b**, The blue and red lines represent impurity scattering (dominant below 40 K) and phonon scattering (dominant above 40 K), respectively. The pink curve is the fit considering both effects. **c**, Coefficients of the phonon (*a*) and impurity scattering (*b*) terms obtained from the fits combining both scattering effects.

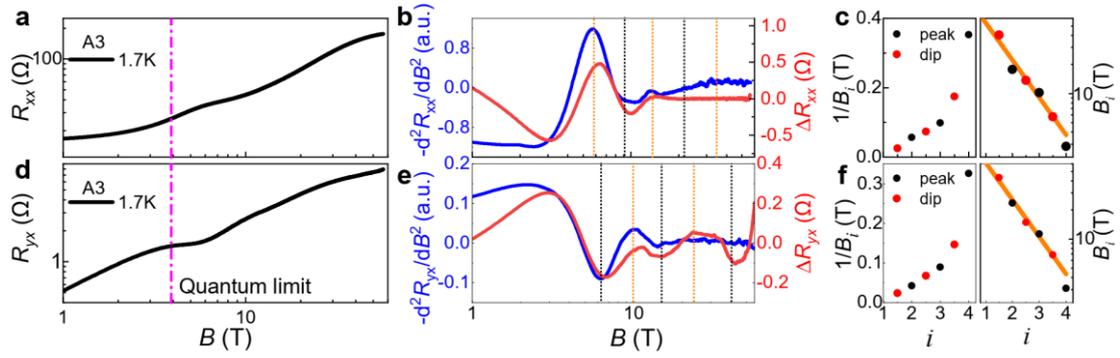

**Extended Data Fig. 5 | Log-periodic oscillations in MR and Hall behavior. a,** Transverse MR of sample A3 at 1.7 K measured in pulsed magnetic fields. **b,** Oscillatory component extracted by background subtraction (red line) and by taking the second derivative (blue line). **c,** Index plot of the MR oscillations. **d-f,** Log-periodic oscillations observed in the Hall resistance of sample A3 at 1.7 K.

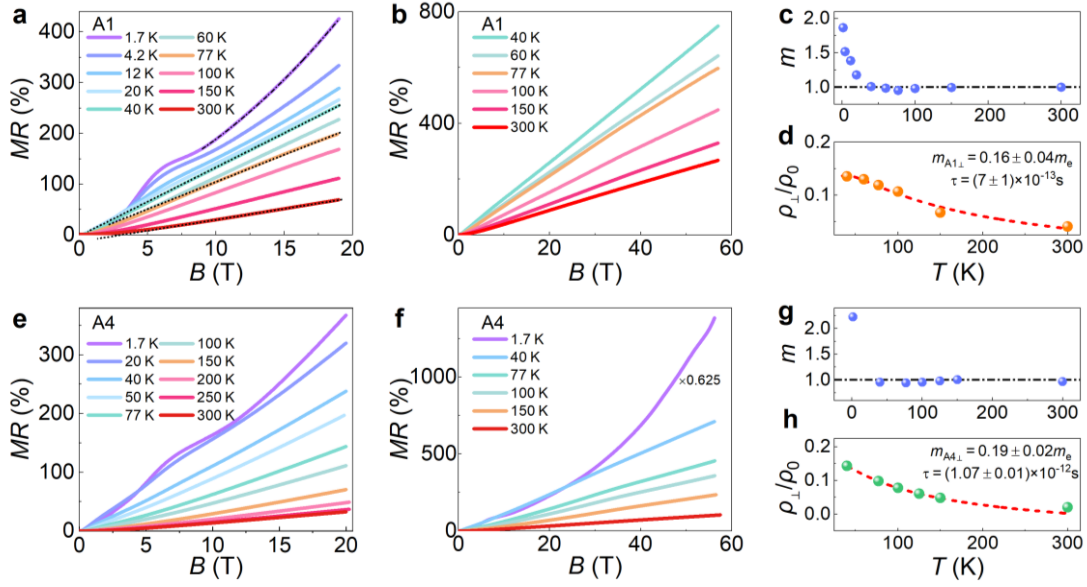

**Extended Data Fig. 6 | Quantum LMR in additional samples. a,** As the temperature rises from 1.7 to 300 K, the large-field MR transitions from a quadratic to a linear dependence on the magnetic field. The black dashed lines serve as guides for the eye. **b,** Robust LMR is observed in sample A1 under high magnetic fields. **c,** Temperature-dependent $m$ obtained by fitting the MR data with MR~ $B^m$. **d,** The transverse LMR slopes show an inverse relationship with temperature. The red dashed line denotes the fit using the phonon-scattering-induced LMR model, and the parameters agree with previous reports for the system. **e-h**, Data from another sample (A4) display consistent results.

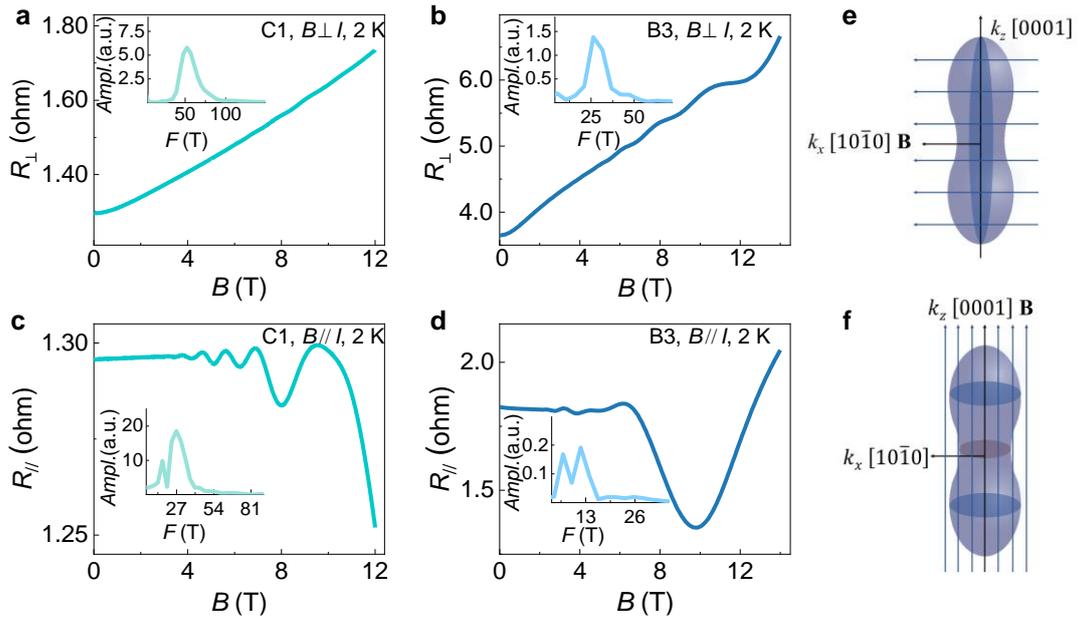

**Extended Data Fig. 7 | SdH oscillations in Te samples with higher carrier densities.**

**a,** MR behavior with SdH oscillations and FFT result (inset) of sample C1 in the transverse configuration. **b,** MR oscillations of sample B3 in the longitudinal configuration. **c, d**, Data from another sample B3, in both configurations. **e** and **f,** Schematics of the anisotropic Fermi surface.